\documentclass[aps,preprint,showpacs,superscriptaddress,groupedaddress]{revtex4}  
\usepackage{graphicx}  
\usepackage{dcolumn}   
\usepackage{bm}        
\usepackage{amssymb}   
\usepackage{amsmath}  

\begin{document}

\title{Vacuum of the Standard Model and its Cosmological Aspects}
\author{Ahmad Mohamadnejad}
\altaffiliation {a.mohamadnejad@ut.ac.ir}
\affiliation{Young Researchers and Elite Club, Islamshahr Branch, Islamic Azad University, Islamshahr 3314767653, Iran}
\date{\today}

\begin{abstract}
Recently, we have proposed a definition for the vacuum and suggest a mechanism for symmetry breaking. In this mechanism extra massless fields, vacuum fields, arise. We apply our method to the standard model of particle physics and obtain vacuum sector of this model. Vacuum field of the standard model is a four-vector with electromagnetic nature. This field is invariant under $ U(1)_{em} $ gauge transformation. We speculate that it could be responsible for many cosmological phenomena like inflation, dark energy and cosmic magnetic fields.
\end{abstract}

\maketitle


Standard model of particle physics is a highly successful model classifying all known elementary particles and describing three fundamental forces (the electromagnetic, weak, and strong interactions). This model is based on the local gauge group $ SU(3)_{C} \otimes SU(2)_{L} \otimes U(1)_{Y} $.
Weinberg \cite{Weinberg} and Salam \cite{Salam} united Higgs mechanism \cite{Higgs,Englert,Guralnik} with Glashow's electroweak interaction \cite{Glashow}, and took a big step towards constructing the standard model. Later, quarks and their interactions were added to the model, giving it its final form.

In this paper, instead of Higgs mechanism, we incorporated our mechanism for symmetry breaking \cite{Mohamadnejad} into the standard model and obtain the vacuum sector of the theory. We begin with this vacuum condition that matter sources of the gauge field equations should be zero. For spinor fields, matter source $ J^{a}_{\mu} $ takes the form
\begin{equation}
J^{a}_{\mu} = g \bar{\psi} T^{a} \gamma_{\mu} \psi .  \label{eq01}
\end{equation}
where $ g $ is the charge of the gauge theory, $ \gamma_{\mu} $ are the Dirac matrices, and $ T^{a} $ are the generators of the gauge group.
Vacuum condition $ J^{a}_{\mu} = 0 $ implies that $ \psi^{V} = 0 $. Therefore, vacuum must be empty of spinor fields (leptons and quarks). We also ignore gluons;
their presence in the vacuum is not permitted because of color confinement. So, the remaining Lagrangian of the standard model is
\begin{eqnarray}
L &=& \frac{1}{2} (D_{\mu} \phi)^{\dagger} (D^{\mu} \phi) - V(\phi^{\dagger} \phi) \nonumber\\
  && - \frac{1}{4} X_{\mu\nu} X^{\mu\nu} - \frac{1}{4} \textbf{W}_{\mu\nu} . \textbf{W}^{\mu\nu}  .  \label{eq02}
\end{eqnarray}
where
\begin{eqnarray}
D_{\mu} \phi  &=& \partial_{\mu} \phi  - ig  W^{i}_{\mu} \frac{\sigma^{i}}{2} \phi  - i g^{'}  X_{\mu} \frac{I}{2} \phi   , \nonumber\\
X_{\mu\nu} &=& \partial_{\mu} X_{\nu} - \partial_{\nu} X_{\mu}   , \nonumber\\
\textbf{W}_{\mu\nu} &=&\partial_{\mu} \textbf{W}_{\nu} - \partial_{\nu} \textbf{W}_{\mu} + g \textbf{W}_{\mu} \times \textbf{W}_{\nu}  , \nonumber\\
V(\phi^{\dagger} \phi) &=&  \frac{\lambda}{4} (\phi^{\dagger} \phi  -  \nu^{2})^{2} , \quad \quad \lambda \, , \, \nu > 0  . \label{eq03}
\end{eqnarray}
The Higgs doublet $ \phi $ can be parameterized:
\begin{equation}
\phi = H \Psi e^{-i\gamma} , \label{eq04}
\end{equation}
with
\begin{equation}
\Psi =
\begin{pmatrix}
\cos{\frac{\alpha}{2}} e^{- i \beta} \\
\thickspace \sin{\frac{\alpha}{2}}
\end{pmatrix}
, \quad \Psi^{\dagger} \Psi = 1 ,  \label{eq05}
\end{equation}
where $ H $, $ \alpha $, $ \beta $, and $ \gamma $ are real fields.

The following conditions minimize the scalar part of the Hamiltonian density ($ H^{S}_{min} = 0 $) and also makes the (scalar) matter source of the gauge fields equal to zero:
\begin{eqnarray}
(D_{\mu} \phi) &=& (D_{\mu} \phi)^{\dagger} = 0 , \label{eq06}  \\
\phi^{\dagger} \phi  &=&  \nu^{2} \, \, \Rightarrow \, H^{V} = \nu .  \label{eq07}
\end{eqnarray}
Therefore
\begin{equation}
\phi^{V} =  \nu \Psi e^{-i\gamma} ,   \label{eq08}
\end{equation}
and thus, according to eq. (\ref{eq06})
\begin{eqnarray}
(\partial_{\mu} \Psi  - ig  W^{i}_{\mu} \frac{\sigma^{i}}{2} \Psi  - i (\partial_{\mu} \gamma + \frac{g^{'}}{2} X_{\mu} ) \Psi)  \Psi^{\dagger} &=& 0 ,  \label{eq09} \\
 \Psi (\partial_{\mu}  \Psi^{\dagger} + ig  W^{i}_{\mu}  \Psi^{\dagger} \frac{\sigma^{i}}{2}  + i (\partial_{\mu} \gamma + \frac{g^{'}}{2} X_{\mu}) \Psi^{\dagger})  &=& 0   .  \label{eq10}
\end{eqnarray}
The sum of eq. (\ref{eq09}) and (\ref{eq10}) yields
\begin{equation}
\partial_{\mu} M  - ig  W^{i}_{\mu} [\frac{\sigma^{i}}{2} , M] =0   ,   \label{eq11}
\end{equation}
where
\begin{equation}
M =  \Psi  \Psi^{\dagger} = \frac{I}{2} + \textbf{n} . \bm{\sigma}   ,   \label{eq12}
\end{equation}
and
\begin{equation}
\textbf{n} =
\begin{pmatrix}
\sin{\alpha} \, \cos{\beta} \\
\sin{\alpha} \, \sin{\beta} \\
\thickspace \cos{\alpha}
\end{pmatrix}
.  \label{eq13}
\end{equation}
Considering eq. (\ref{eq11}) and eq. (\ref{eq12}), we have
\begin{eqnarray}
&& \partial_{\mu} \textbf{n}  + g  \textbf{W}_{\mu} \times \textbf{n} = 0 ,  \label{eq14}   \\
&& \Rightarrow \textbf{W}^{V}_{\mu} = W^{V}_{\mu}  \textbf{n} + \frac{1}{g} \partial_{\mu} \textbf{n} \times \textbf{n} ,  \label{eq15}
\end{eqnarray}
where $  W^{V}_{\mu} = \textbf{W}^{V}_{\mu} .\textbf{n}  $.
Now eq. (\ref{eq09}) implies
\begin{equation}
 \Psi^{\dagger} (\partial_{\mu} \Psi  - ig  \textbf{W}^{V}_{\mu} . \frac{\bm{\sigma}}{2} \Psi  - i (\partial_{\mu} \gamma + \frac{g^{'}}{2} X_{\mu} ) \Psi) = 0  , \label{eq16}
\end{equation}
and according to
\begin{eqnarray}
\Psi^{\dagger} \bm{\sigma} \Psi &=&  \textbf{n}   , \, \,   \Psi^{\dagger} \Psi = 1 , \label{eq17}   \\
and \quad \Psi^{\dagger} \partial_{\mu} \Psi &=& - i \cos^{2}{\frac{\alpha}{2}} \partial_{\mu} \beta , \label{eq18}
\end{eqnarray}
we get
\begin{equation}
X^{V}_{\mu} = - \frac{g}{g^{'}} W^{V}_{\mu} -  \frac{2}{g^{'}} \cos^{2}{\frac{\alpha}{2}} \partial_{\mu} \beta - \frac{2}{g^{'}}  \partial_{\mu} \gamma . \label{eq19}
\end{equation}
Substituting eqs. (\ref{eq08}),  (\ref{eq15}), and (\ref{eq19}) in the Lagrangian (\ref{eq02}), the vacuum Lagrangian will be
\begin{equation}
L_{V} = - \frac{1}{4\sin^{2}{\theta_{W}}} (W^{V}_{\mu\nu})^{2} , \label{eq20}
\end{equation}
where
\begin{equation}
W^{V}_{\mu\nu} = \partial_{\mu} W^{V}_{\nu} - \partial_{\nu} W^{V}_{\mu} - \frac{\sin{\alpha}}{g}  (\partial_{\mu} \alpha \, \partial_{\nu} \beta - \partial_{\nu} \alpha \, \partial_{\mu} \beta)   , \label{eq21}
\end{equation}
and $ \theta_{W} $ is the Weinberg angle ($ \sin{\theta_{W}} = \frac{g^{'}}{\sqrt{g^{2} + g^{'2}} } $).

To obtain a more physical Lagrangian, we now make a use of the fact that we have a local symmetry, so may perform independent gauge transformations at each point in space-time. We therefore select a gauge, unitary gauge in which only physical particles appear, so that:
\begin{equation}
\phi \rightarrow e^{-i \frac{\alpha_{1}}{2} \bm{\sigma}. \textbf{m} } e^{-i \frac{\beta_{1}}{2} } \phi =
\begin{pmatrix}
0 \\
H
\end{pmatrix}, \label{eq22}
\end{equation}
where
\begin{equation}
\alpha_{1} = \pi - \alpha,  \, \, \beta_{1} = - 2 \gamma , \, \, \textbf{m} =
\begin{pmatrix}
- \sin{\beta} \\
 \, \, \cos{\beta} \\
\thickspace 0
\end{pmatrix} . \label{eq23}
\end{equation}
For vacuum gauge fields in unitary gauge we have
\begin{eqnarray}
X^{V}_{\mu} &\rightarrow& X^{V}_{\mu} - \frac{1}{g^{'}} \partial_{\mu} \beta_{1}   , \label{eq24}   \\
 (\textbf{W}^{V}_{\mu} . \frac{\bm{\sigma}}{2}) &\rightarrow&  U (\textbf{W}^{V}_{\mu} . \frac{\bm{\sigma}}{2}) U^{\dagger}
- \frac{i}{g} (\partial_{\mu} U) U^{\dagger} , \label{eq25}
\end{eqnarray}
where
\begin{eqnarray}
U = e^{-i \frac{\alpha_{1}}{2} \bm{\sigma}. \textbf{m} } = \cos{\frac{\alpha_{1}}{2}} - i \bm{\sigma}. \textbf{m} \sin{\frac{\alpha_{1}}{2}} . \label{eq26}
\end{eqnarray}
According to eqs. (\ref{eq25}) and (\ref{eq26}):
\begin{eqnarray}
&& \textbf{W}^{V}_{\mu} \rightarrow (\textbf{W}^{V}_{\mu} . \textbf{m})  \textbf{m} - \sin{\alpha_{1}}  (\textbf{W}^{V}_{\mu} \times \textbf{m})  \nonumber\\
&& +  \cos{\alpha_{1}}  \textbf{m} \times (\textbf{W}^{V}_{\mu} \times \textbf{m})
 - \frac{1}{g} (\partial_{\mu} \alpha_{1}) \textbf{m}  +  \frac{1}{g} \partial_{\mu} \textbf{m} \times \textbf{m}  \nonumber\\
&& -  \frac{\sin{\alpha_{1}}}{g} \partial_{\mu} \textbf{m} - \frac{\cos{\alpha_{1}}}{g} \partial_{\mu} \textbf{m} \times \textbf{m} . \label{eq27}
\end{eqnarray}
Now it is straightforward, if tedious, to show that in unitary gauge vacuum fields are
\begin{eqnarray}
&& (W^{1}_{\mu})^{V} =(W^{2}_{\mu})^{V}= 0, \nonumber\\
&& (W^{3}_{\mu})^{V}= - W^{V}_{\mu}-  \frac{2}{g} \cos^{2}{\frac{\alpha}{2}} \partial_{\mu} \beta   , \label{eq28}
\end{eqnarray}
and
\begin{equation}
X^{V}_{\mu} =   - \frac{g}{g^{'}} W^{V}_{\mu} -  \frac{2}{g^{'}} \cos^{2}{\frac{\alpha}{2}} \partial_{\mu} \beta . \label{eq29}
\end{equation}
According to the following definitions
\begin{eqnarray}
(W^{\pm}_{\mu})^{V} &=& \frac{1}{\sqrt{2}} ((W^{1}_{\mu})^{V} \mp (W^{2}_{\mu})^{V}) ,  \nonumber\\
Z^{V}_{\mu} &=&  \frac{g (W^{3}_{\mu})^{V} - g^{'} X^{V}_{\mu}}{\sqrt{g^{2} + g^{'2}}} , \nonumber\\
A^{V}_{\mu} &=&  \frac{g^{'} (W^{3}_{\mu})^{V} + g X^{V}_{\mu}}{\sqrt{g^{2} + g^{'2}}} , \label{eq30}
\end{eqnarray}
vacuum fields in unitary gauge are
\begin{eqnarray}
&& (W^{\pm}_{\mu})^{V} =  Z^{V}_{\mu} = 0,  \label{eq31} \\
&&  A^{V}_{\mu} = - \frac{1}{\sin{\theta_{W}}} (W^{V}_{\mu} + \frac{2}{g} \cos^{2}{\frac{\alpha}{2}} \partial_{\mu} \beta ) . \label{eq32}
\end{eqnarray}
Therefore, the nature of the vacuum field $ V_{\mu}= A^{V}_{\mu} $ is electromagnetic.
and massive gauge bosons do not contribute in vacuum.

Using eq. (\ref{eq18}) the vacuum field will be
\begin{equation}
V_{\mu}= A_{\mu} + B_{\mu} , \label{eq34}
\end{equation}
where
\begin{eqnarray}
A_{\mu} &=&  - \frac{1}{\sin{\theta_{W}}} W^{V}_{\mu} ,  \label{eq35} \\
B_{\mu} &=&  \frac{1}{i e} (\Psi^{\dagger} (\partial_{\mu} \Psi) -  (\partial_{\mu} \Psi^{\dagger})\Psi )  , \label{eq36}
\end{eqnarray}
where $ e = g \sin{\theta_{W}} $ is the proton charge.
Remarkably  $ V_{\mu} $ itself is invariant under $ U(1)_{em} $ gauge transformations:
\begin{eqnarray}
\Psi &\rightarrow&  e^{- \frac{ i}{2} \omega} \Psi ,  \label{eq37} \\
A_{\mu} &\rightarrow&  A_{\mu} + \frac{1}{e} \partial_{\mu} \omega  . \label{eq38}
\end{eqnarray}
Therefore adding a potential term to the vacuum Lagrangian does not violate gauge invariance:
\begin{equation}
L= - \frac{1}{4} V_{\mu \nu} V^{\mu \nu} - V(V_{\mu}V^{\mu}) , \label{eq39}
\end{equation}
where $ V_{\mu \nu} = \partial_{\mu}V_{\nu} - \partial_{\nu}V_{\mu} $, and $ V(V_{\mu}V^{\mu}) $ is a gauge invariant potential term. If we minimally  couple the vacuum field to the gravity we get this action:
\begin{equation}
s= \int d^{4}x \sqrt{-g} [\frac{-1}{16 \pi G} R - \frac{1}{4} V_{\mu \nu} V^{\mu \nu} - V(V_{\mu}V^{\mu})]. \label{eq40}
\end{equation}
where $ g = det(g_{\mu \nu }) $ is the determinant of the metric tensor matrix, $ G $ is the gravitational constant, and
 $ R $ is the Ricci scalar.
This kind of action is used in a model \cite{Ford} in which inflation \cite{Guth,Linde} is driven by a vector field rather than a scalar field. The big difference here is that unlike \cite{Ford}, our formalism enjoys gauge invariance. Furthermore, vacuum field $ V_{\mu} $ is composed of another vector field $ A_{\mu} $, and a scalar doublet $ \Psi $, and this may change the results. Indeed, vector fields coupled to gravity is proposed by many authors, for example see \cite{Mukhanov,Mota,Picon}, in order to describe both inflation and late time accelerating expansion of the universe \cite{D1,D2} related to the dark energy.
But the nature of the vector field is unknown and they suffer from violating gauge invariance, while in our formulation the nature of the vector field is electromagnetic and its interaction with other standard model particles is not unknown.

Now we consider vacuum Lagrangian without potential term:
\begin{equation}
L= - \frac{1}{4} V_{\mu \nu} V^{\mu \nu} = - \frac{1}{4} (F_{\mu \nu}  + B_{\mu \nu} )(F^{\mu \nu}  + B^{\mu \nu} ) , \label{eq41}
\end{equation}
where
\begin{eqnarray}
F_{\mu \nu} &=&  \partial_{\mu}A_{\nu} - \partial_{\nu}A_{\mu}  ,  \label{eq43} \\
B_{\mu \nu} &=&   \frac{2}{i e} (\partial_{\mu}\Psi^{\dagger} \partial_{\nu} \Psi -  \partial_{\nu}\Psi^{\dagger} \partial_{\mu} \Psi) . \label{eq44}
\end{eqnarray}
The Euler-Lagrange equation of the vacuum Lagrangian by variation with respect to $ A_{\mu} $ yields
\begin{equation}
\partial_{\mu} V^{\mu \nu}  = \partial_{\mu} F^{\mu \nu}  + \partial_{\mu} B^{\mu \nu} = 0  , \label{eq45}
\end{equation}
and variation with respect to $ \Psi $ does not lead to a new equation. Therefore in the absence of the potential term, the scalar doublet $ \Psi $ is not a dynamical field.
However, it carries energy and momentum.
Note that $ \Psi $ is a charged field and it couples with $ A_{\mu} $ and charged particles.
According to the eq. (\ref{eq08}) this field is nothing but the orientation of the Higgs doublet vacuum state in every space-time points. If we consider one vacuum state for all space-time, then $ \Psi $ will be a constant and its contribution to the vacuum Lagrangian disappears. However, we suppose that it could be a function and because it is not a dynamical field we assume that it is not time dependent and it simply  shows the orientation of the Higgs doublet vacuum state in every space point. In traditional approach $ \Psi $ is a constant and orientation of the vacuum state is the same for all space. This approach works well in particle physics. However, we consider a possibility that it can change at least on cosmological scales. If  $ \Psi $ is only a function of space on large scales, then according to eq. (\ref{eq44}) it can only generate large scale magnetic fields or cosmic magnetic fields.
Because of homogeneity and isotropy of space we expect that on large scales $ \Psi $ should be randomly oriented and thus it generates randomly oriented cosmic magnetic fields with almost constant strength.
Indeed, cosmological observations show that magnetic fields in the $ \mu G $ range are pervading; for a review see \cite{G}. All galaxies and clusters of galaxies
carry magnetic fields that are large and extensive with unknown origin.
Even more mysterious is the ubiquity of these fields and the uniformity of their strength.
Though strong homogeneous fields are ruled out by the uniformity of the cosmic
background radiation, large domains with uniform fields are possible.
The origin of cosmic magnetic field is an outstanding problem in modern cosmology which according to our proposal it could be related to the vacuum field $ \Psi $.
If we relate the vacuum field $ A_{\mu} $ to the cosmic microwave background (CMB) radiation which fills vacuum uniformly and associate $ \Psi $ to the cosmic magnetic fields, using equipartition theorem, we conclude that the energy density of the magnetic field should be the same order as CMB energy density. It is because $ A_{\mu} $ has two degrees of freedom (associated with two polarization of the photon) and $ \Psi $ has also two degrees of freedom ($ \alpha $ and $ \beta $).
Using energy density of CMB, $ \rho_{r} = 4.64 \times 10^{-31} \, Kg.m^{-3} $, equipartition theorem yields  $  B_{avg} \simeq 3 \mu G  $ where $  B_{avg}  $ is the averaged magnetic field . This estimation agrees fairly well with the astronomical observations. Assuming Robertson-Walker metric, if the orientation of $ \Psi $ does not change in co-moving Robertson-Walker coordinate, then according to eq. (\ref{eq44}) we expect that
\begin{eqnarray}
B &=&  \frac{a^2_{0}}{a^2} B_{0} ,  \nonumber \\
 &\Rightarrow&  \rho_{B} = \frac{1}{2} B^{2} = \frac{a^4_{0}}{a^4} \rho_{{B_{0}}}. \label{eq46}
\end{eqnarray}
where $  a  $ is the Robertson-Walker scale factor.
Therefore, magnetic energy density $ \rho_{B} $ behaves as the same as radiation energy density $ \rho_{r} $ as the universe expands and if they are in equilibrium now, they will remain in equilibrium.

Electromagnetic effects on cosmological scales should not be a surprise. Apart from gravity, only electromagnetism acts as a long-range interaction which could become relevant on cosmological scales. We show that vacuum sector of the standard model has a electromagnetic nature with a vector field and a scalar field. The scalar field is a charged field and it is possible that due to some asymmetry, like matter-antimatter asymmetry, it generates uniform electric charge in the universe. Therefore, even the assumption of exact electric neutrality of the universe on large scales is debatable.
If this is the case, the attractive gravitational force is not the only relevant force on large scales and electromagnetic force, which could be repulsive, can also play a crucial role.

This work is supported financially by the Young Researchers and Elite Club of Islamshahr Branch of Islamic Azad University.

\end{document}